\documentclass[12pt]{article}

\usepackage{amsfonts,amssymb,amsmath}
\usepackage{tikz}

\renewcommand{\ker}{{\rm ker}}
\newcommand{\im }{{\rm im}}
\newcommand{\coker}{{\rm coker}}
\newcommand{\Hom}{{\rm Hom}}
\newcommand{\ra}{{\rightarrow}}
\newcommand{\cG}{{\mathbb G}}
\newcommand{\ZZ}{{\mathbb Z}}
\newcommand{\RR}{{\mathbb R}}

\newcommand{\Aut}{{\rm Aut}}
\newcommand{\tf}{{\widetilde f}}

\newcommand{\frg}{{\mathfrak g}}
\newcommand{\frh}{{\mathfrak h}}
\newcommand{\dt}{{\bar t}}
\newcommand{\da}{{\bar\alpha}}

\newcommand{\sa}{{\mathsf a}}
\newcommand{\RZ}{{\mathbb{R}/\mathbb{Z}}}
\renewcommand{\sb}{{\mathsf b}}
\newcommand{\cL}{{\mathcal L}}
\newcommand{\cA}{{\mathcal A}}
\newcommand{\hPi}{{\hat \Pi}}
\newcommand{\bbH}{{\mathbb H}}

\usetikzlibrary{decorations.markings}
\usetikzlibrary{positioning}
\usetikzlibrary{shadings}

\author{Anton Kapustin \\ {\it California Institute of Technology, Pasadena, CA} \and Ryan Thorngren \\ {\it University of California, Berkeley, CA}}

\title{Higher symmetry and gapped phases of gauge theories}

\begin{document}

\tikzset{->-/.style={decoration={
  markings,
  mark=at position #1 with {\arrow[scale=2]{>}}},postaction={decorate}}}

\titlepage
\maketitle

\abstract{We study topological field theory describing gapped phases of gauge theories where the gauge symmetry is partially Higgsed and partially confined. The TQFT can be formulated both in the continuum and on the lattice and generalizes Dijkgraaf-Witten theory by replacing a finite group by a finite 2-group. The basic field in this TQFT is a 2-connection on a principal 2-bundle. We classify topological actions for such theories as well as loop and surface observables. When the topological action is trivial, the TQFT is related to a Dijkgraaf-Witten theory  by electric-magnetic duality, but in general it is distinct. We propose the existence of new phases of matter protected by higher symmetry.}

\section{Introduction and summary}

Gapped phases of matter are described at long distances by unitary Topological Quantum Field Theories. Thus it is of great interest to try to classify unitary TQFTs, at least in space-time dimensions up to four. Since gauge interactions are ubiquitous in nature, it is particularly interesting to identify TQFTs which describe gapped phases of gauge theories. In the case when the microscopic gauge group is Higgsed down to a finite group $G$, a complete classification (at least if no further global symmetries are postulated) has been given by Dijkgraaf and Witten \cite{DW}. Namely, topological actions in $d$ dimensions are classified by degree-$d$ cohomology classes for $G$ with coefficients in $U(1)$. But DW theories do not exhaust all possibilities.  For example, in 3d there are Chern-Simons theories, which are more general than 3d DW theories. In this paper we study another class of TQFTs which exist in all dimensions and are more general than DW theories. These TQFTs involve both 1-form and 2-form gauge fields, as well as 0-form and 1-form gauge symmetries. 1-form gauge fields take values in the Lie algebra of a Lie group $G$, while 2-form fields take values in the Lie algebra of another Lie group $H$.  There is a homomorphism $t$ from $H$ to $G$ and an action $\alpha$ of $G$ on $H$ which enable one to write down a consistent set of transformation rules. The quadruple $\cG=(G,H,t,\alpha)$ is what is known as a 2-group (see the next section). It has been argued in \cite{GK} that such TQFTs describe massive phases of gauge theories where the microscopic gauge group is partially confined and partially Higgsed. More precisely, $G$ is the part of the  microscopic gauge group which is not Higgsed, and the image of $t$ is the the confined subgroup of $G$. In this paper we analyze such TQFTs along the lines of \cite{DW}. 

In the case when $H$ is trivial, our TQFT reduces to a DW theory. The other extreme is when $H=G$ and $t$ is a surjection. In this case the whole gauge group is confined, but the theory may have a nontrivial ``magnetic gauge group" given by $\ker\ t$. Such theories have been recently analyzed in \cite{KT1,T} (see also \cite{Quinn} and section 5 of \cite{FHLT}) . The present paper generalizes both \cite{DW} and \cite{KT1,T}.

Here is a brief summary of our results. We show that TQFTs which describe phases with both Higgs effect and confinement depend on the following data:  a finite (possibly non-Abelian) group $\Pi_1$, a finite abelian group $\Pi_2$, an action of $\Pi_1$ on $\Pi_2$, and a degree-3 group cohomology class for $\Pi_1$ with coefficients in $\Pi_2$. Here $\Pi_1$ is interpreted as the low-energy ``electric'' gauge group while $\Pi_2$ is the ``magnetic'' gauge group. For every such quadruple and every $d>1$ we construct a lattice model in $d$ dimensions generalizing the DW model. It describes a lattice gauge field with gauge group $\Pi_1$ coupled to a lattice 2-form with values in $\Pi_2$. We classify possible action for such models in dimensions $2$, $3$ and $4$. In dimension $2$ the theory is equivalent to the DW theory for $\Pi_1$. In dimension $3$ the action depends on two parameters: a degree-3 cohomology class for $\Pi_1$ with coefficients in $U(1)$ and a degree-1 cohomology class for $\Pi_1$ with coefficients in $\hPi_2=\Hom(\Pi_1,U(1))$. In dimension $4$ the action depends on three parameters: a degree-4 cohomology class for $\Pi_1$ with coefficients in $U(1)$, a degree-2 cohomology class for $\Pi_1$ with coefficients in $\hPi_2$, and a quadratic function on $\Pi_2$ with values in $U(1)$. We show that in dimension $3$ the lattice 2-form can be dualized to a lattice scalar with values in $\hPi_2$. In dimension $4$ the lattice 2-form can be dualized to a lattice gauge field provided only the first two terms in the action are nonzero. We also classify TQFT observables, including  surface observables measuring the flux of the 2-form. 

Recently DW theories found a new application: it has been argued that topological actions for a DW theory with gauge group $G$ classify gapped phases of matter with global symmetry $G$ and no long-range entanglement \cite{SPT, LevinGu, SPT2}. Such gapped phases are called symmetry-protected phases. We propose that there are more general phases which are protected not by a symmetry group, but by a symmetry 2-group. While an ultralocal internal symmetry group acts on degrees of freedom living on sites of a lattice, a ultralocal 2-symmetry acts on both site and link variables. The groups $\Pi_1$ and $\Pi_2$ describe symmetry transformations which live on sites and links, respectively.  Gapped phases protected by a 2-group symmetry $\cG$ are classified by topological actions for a TQFT based on $\cG$. In the special case when only the magnetic gauge group $\Pi_2$ is nontrivial (the case considered in detail in \cite{KT1}) it appears that the 2-group TQFT is equivalent to a special case of the Walker-Wang TQFT \cite{WW, KBS}. 

More generally, as explained below, in $d$ space-time dimensions one can contemplate symmetry $p$-groups with $p\leq d$, and accordingly one can have gapped phases with short-range entanglement protected by such higher symmetry.

A.K. would like to thank Dan Freed, Sergei Gukov, Michael Hopkins, Nathan Seiberg, Yuji Tachikawa, and Constantin Teleman for discussions. R.T. would like to thank Scott Carnahan, Curt von Keyserlingk, Evan Jenkins, Alex Rasmussen, David Roberts, and Urs Schreiber for discussions.
This work was supported in part by the DOE grant DE-FG02-92ER40701 and by the National Science Foundation under Grant No. PHYS-1066293 and the hospitality of the
Aspen Center for Physics.

\section{2-groups}

We begin by recalling the notion of a 2-group \cite{BaezLauda} and its physical interpretation \cite{GK}. The most concise definition uses the language of higher categories.
A (weak) 2-group is a weak 2-category with a single object and such that  all 1-morphisms are weakly-invertible and 2-morphisms are invertible. This is analogous to the definition of a group as a category with a single object and such that all morphisms are invertible. 

The most pedestrian definition goes as follows. A 2-group is a quadruple $\cG=(G,H,t,\alpha)$ where $G$ and $H$ are groups, $t:H\ra G$ is a group homomorphism, and $\alpha:G\ra \Aut(H)$ is an action of $G$ on $H$ such that the following two identities hold:
\begin{equation}\label{crossedmodule}
t(\alpha(g)(h))=g t(h) g^{-1},\quad \alpha(t(h))(h')=h h' h^{-1}. 
\end{equation}
A 2-group defined in this way is also known as a crossed module. The relation between the two definitions is this: $G$ is the set of 1-morphisms, $H$ is the set of 2-morphisms from the identity 1-morphism to all other 1-morphisms, the map $t$ assigns to a 2-morphism its target 1-morphism. 

Let us give a few examples of 2-groups. One simple class of examples is obtained by taking $G$ to be a connected simple Lie group, $t:H\ra G$ to be a covering map, and $\alpha$ to be given by conjugation: $h\mapsto \widetilde g h \widetilde g^{-1}$, where $\widetilde g=t^{-1} g$. Note that while $t^{-1}$ is multivalued, different choices are different only by elements of the center of $H$, and therefore $\alpha$ is well-defined. A nontrivial finite example is provided by $H=G=\ZZ_4$, $t(n)=2n$, and $\alpha(n)(m)=(-1)^n m$.

We are interested in the case when $G$ and $H$ are Lie groups, and the maps $t$ and $\alpha$ are smooth; such a 2-group is called a Lie 2-group. The physical meaning of these data is as follows: $G$ is the subgroup of the microscopic gauge group which is not broken by the Higgs effect, $\im\ t\subset G$ is the part of $G$ which is confined due to monopole condensation (this subgroup is normal, as a consequences of the identities (\ref{crossedmodule}), $\coker\ t=G/\im\ t$ is the low-energy gauge group, $\ker\ t$ is the group whose elements label conserved magnetic  fluxes (it is a subgroup of the center of $H$; we will call it the magnetic gauge group). In the above finite example, the unbroken gauge group is $\ZZ_4$, its $\ZZ_2$ subgroup is confined, so the low-energy electric gauge group is $\ZZ_2$. The magnetic gauge group is also $\ZZ_2$, and the electric gauge group acts on the magnetic fluxes by ``charge conjugation". 

For groups we have an obvious notion of an isomorphism; for 2-groups a better notion is that of equivalence (of 2-categories). This notion does not preserve the sets $G$ and $H$, it only preserves the sets $\coker\ t$ and $\ker\ t$. This is very reasonable from the physical point of view: representations of $\coker\ t$ describe electric sources visible at long distances, elements of $\ker\ t$ describe magnetic sources, but $G$ and $H$ themselves are not observable at long distances. We will see that for any 2-group $\cG$ there is a TQFT which depends only on the equivalence class of $\cG$. 

It is thus natural to ask what sort of data describe the equivalence class of a 2-group. It turns out it is a quadruple $(\Pi_1,\Pi_2,\alpha,\beta)$, where $\Pi_1=\coker\ t$ is a group, $\Pi_2=\ker\ t$ is an abelian group, $\alpha$ is an action of $\Pi_1$ on $\Pi_2$ (it descends from the action of $G$ on $H$, so we denote it by the same letter), and $\beta$ is an element of $H^3(B\Pi_1,\Pi_2)$. Here and below $BG$ denotes the classifying space of principal $G$-bundles. If the group $G$ is finite, it is defined up to homotopy by the condition $\pi_n(BG)=0$ for $n>1$ and  $\pi_1(BG)=G$ \cite{Hatcher}. By definition, there is a principal $G$-bundle $EG$ with base $BG$ such that the total space of $EG$ is contractible. Group cohomology of $G$ is defined in terms of $EG$. Namely, if $G$ acts on an abelian group $M$, we have an associated flat bundle over $BG$ with fiber $M$, and $H^d(BG,M)$ denotes the degree-$d$ cohomology of $BG$ with coefficients in this bundle.

The definition of $\alpha: \Pi_1\ra \Aut(\Pi_2)$ should be clear: one first checks that the action of $G$ on $H$ maps $\ker\ t$ to itself, and then that elements of $\im\ t$ act trivially on $\ker\ t$. This implies that $\coker\ t$ acts on $\ker\ t$. Physically, this is also quite natural: the residual gauge group may act on the magnetic flux, like in the case of the gauge group $O(2n)$, where charge conjugation is part of the gauge group and flips the sign of the magnetic flux. 

The definition of $\beta$ is less obvious. Mathematically, one uses the fact \cite{Brown} that elements of $H^3(B\Pi_1,\Pi_2)$ can be interpreted as equivalence classes of double extensions of $\Pi_1$ by $\Pi_2$, with a fixed action of $\Pi_1$ on $\Pi_2$. That is, these are equivalence classes of four-term exact sequences
$$
1\ra \Pi_2\ra H'\ra G'\ra \Pi_1\ra 1,
$$
where $(G',H',t',\alpha')$ is a crossed module with $\coker\ t'=\Pi_1$,  $\ker\ t'=\Pi_2$, and with $\alpha'$ inducing our chosen action of $\Pi_1$ on $\Pi_2$. 
To get $\beta$ we simply take the tautological sequence
$$
1\ra \ker\ t\ra H\ra G\ra \coker\ t \ra 1
$$
We will see that the TQFT depends on $(\Pi_1,\Pi_2,\alpha,\beta)$ and nothing else. However, the continuum action is more conveniently written using the data $(G,H,t,\alpha)$. This is because we normally assume that $\Pi_1$ and $\Pi_2$ are finite (to ensure that the low-energy theory is gapped), but to write down a continuum action it is convenient to work with fields which take values in the Lie algebras of $G$ and $H$. 

The most basic example of a group arising from topology is the fundamental group of a (pointed connected) topological space. This group encodes the homotopy 1-type of a space. Similarly, there is a 2-group associated to any pointed connected topological space which encodes its homotopy 2-type. This 2-group has $G=\pi_1(X_1,x_0)$, $H=\pi_2(X,X_1,x_0)$, where $X_1$ is a 1-skeleton of $X$, and $x_0$ is the marked point. We will call it the fundamental 2-group of $X$. The homotopy equivalence class of these data is encoded by the quadruple $(\pi_1(X,x_0),\pi_2(X,x_0),\alpha,\beta)$, where $\alpha$ is the usual monodromy action of $\pi_1$ on $\pi_2$ and $\beta\in H^3(B\pi_1,\pi_2)$ is the so-called Postnikov invariant of $X$.

\section{2-gauge theory in the continuum}

The path-integral of the DW theory is constructed as an integral over the moduli space of flat connections with a structure group $G$. Similarly, we can construct a TQFT whose path-integral is an integral over the moduli space of flat 2-connections with a structure 2-group $\cG$. But first we need to define the notion of a flat 2-connection on a manifold $X$. 

A flat connection for a group $G$ can be defined as a homomorphism from $\pi_1(X)$ to $G$. Similarly, a flat 2-connection for a 2-group $\cG$ can be defined as a ``2-homomorphism'' from the fundamental 2-group of $X$ to $\cG$. However, both of these definitions are not manifestly local and therefore not suitable for constructing a TQFT . A local definition of a flat 2-connection goes as follows \cite{BaezSchreiber, SchreiberWaldorf}.  

Let $\frg$, $\frh$ be the Lie algebras of $G$ and $H$. Let $\dt: \frh\ra\frg$ be the homomorphism of Lie algebras induced by the homomorphism of Lie groups $t:H\ra G$. Since $\ker\ t$ and $\coker\ t$ were assumed to be finite, $\dt$ is an isomorphism. Let $\da: G\ra \Aut(\frh)$ be the action of $G$ on $\frh$ induced by $\alpha$. 
Let $\left\{U_i\right\}$, $i\in I$, be a cover of $X$ such that all overlaps are contractible. A flat 2-connection with structure 2-group $\cG$ is defined by the following data. On each chart $U_i$ one has  a $\frg$-valued 1-form $A_i$, on each $U_{ij}$ one has a $G$-valued function $g_{ij}$ and an $\frh$-valued 1-form $\lambda_{ij}$, and on each $U_{ijk}$ one has an $H$-valued function $h_{ijk}$ so that the following conditions are satisfied:
\begin{itemize}
\item On each $U_{ij}$ one has $A_j=g_{ij} A_i g_{ij}^{-1}+g_{ij} dg_{ij}^{-1}-\dt(\lambda_{ij})$.
\item On each $U_{ijk}$ one has $g_{ik}=t(h_{ijk})g_{jk} g_{ij}$, and
\begin{eqnarray}
h_{ijk}^{-1}\lambda_{ik} h_{ijk} & = & \da (g_{jk})(\lambda_{ij})+\lambda_{jk} -h_{ijk}^{-1} d h_{ijk} \nonumber \\
 & & -h_{ijk}^{-1}\, \dt^{-1}(A_k) h_{ijk} + \dt^{-1} (A_k) \,. \nonumber
\end{eqnarray}
\item On each $U_{ijkl}$ one has $h_{ijl}h_{jkl} \, = h_{ikl} \cdot \alpha(g_{kl})(h_{ijk}).$
\end{itemize}
We can also introduce $\frh$-valued 2-forms $B_i$ on each $U_i$ by $B_i=\dt^{-1} F_{A_i}$, where $f_1=dA+A\wedge A$. Alternatively, we can treat them as independent 2-forms, with suitable gluing conditions on double overlaps, and regard the condition $F_{A_i}=\dt(B_i)$ as an equation of motion. 

Just like in the case of ordinary connections, there is a notion of gauge equivalence of flat 2-connections. 
Let $(A_i,g_{ij},\lambda_{ij},h_{ijk})$ and $(A_i',g_{ij}',\lambda_{ij}',h_{ijk}')$ be a pair of 2-connections. A gauge-equivalence between them is a $G$-valued  function $g_i$ and an $\frh$-valued 1-form $\lambda_i$ on each $U_i$, together with an $H$-valued function $h_{ij}$ on every $U_{ij}$ such that
\begin{itemize}
\item On each $U_i$ one has 
$$
A_i'=g_i A_i g_i^{-1}+g_i dg_i^{-1}-\dt(\lambda_i).
$$
\item On each $U_{ij}$ one has $g_{ij}'=t(h_{ij})g_j g_{ij} g_i^{-1},$ and
\begin{eqnarray}
\lambda_{ij}' & = & h_{ij} \left(\alpha(g_j)(\lambda_{ij})+\lambda_j\right)h_{ij}^{-1}-\alpha(g'_{ij})(\lambda_i) \nonumber \\
 & & +h_{ij}dh_{ij}^{-1}+h_{ij}\dt^{-1}(A_j') h_{ij}^{-1}-\dt^{-1}(A_j')  \,. \nonumber
\end{eqnarray}
\item On each $U_{ijk}$ one has
$$
h_{ijk}'=h_{ik}\alpha(g_k)(h_{ijk})h_{jk}^{-1}\alpha(g_{jk}')(h_{ij}^{-1}).
$$
\end{itemize}
There are also 2-gauge transformations between gauge transformations, see \cite{BaezSchreiber, SchreiberWaldorf, GK} for details.

If 2-forms $B_i$ are regarded as independent variables, then the action in $d$ space-time dimensions has the form
$$
S=\int \langle (f_1-\dt(B)), \wedge b\rangle +S_{top},
$$
where $b$ is a Lagrange multiple $(d-2)$-form with values in $\frg^*$ and $S_{top}$ is a topological action which is invariant under gauge equivalence of flat 2-connections. If we regard $B_i$ as dependent variables, then the action contains only $S_{top}$. Below we will determine the most general form of $S_{top}$ in dimension $4$ and lower.

For now, let us show that the moduli space of flat 2-connections can be reformulated purely in terms of the data $(\Pi_1,\Pi_2,\alpha,\beta)$. First, we use the 1-form gauge transformations to set $A_i=0$ (and consequently $B_i=0$). This ensures that the 1-forms $\lambda_{ij}$ is pure gauge:
$$
\lambda_{ij}=\dt^{-1}(g_{ij} dg_{ij}^{-1}),
$$
and simultaneously that for any two points  $p,p'\in U_{ij}$ we have 
$$g_{ij}^{-1}(p)g_{ij}(p')\in \im\ t.$$
Thus if we project $g_{ij}$ to $\Pi_1=\coker\ t$, the resulting $\Pi_1$-valued function is constant. We denote this function $\sa_{ij}$. 

Next, let us choose a map $s: \Pi_1\ra G$ which is an inverse of the projection $G\ra \Pi_1$. In general, we cannot choose $s$ to be a homomorphism, but we can always choose it so that $s(x)^{-1}=s(x^{-1})$ for all $x\in\Pi_1$. By definition, $s(\sa_{ij}) g_{ij}^{-1}$ is a function on $U_{ij}$ which takes values in $\im\ t$. Thus there exists an $H$-valued function $h_{ij}$ such that $t(h_{ij})=s(\sa_{ij}) g_{ij}^{-1}$. Performing another gauge transformation, now with
$\lambda_i=0$, $g_i=1$ and $h_{ij}$ chosen as above, we bring the data defining a 2-connection to the form 
$$
A_i=0,\quad \lambda_{ij}=0,\quad g_{ij}=s(\sa_{ij}).
$$
In addition, the functions $h_{ijk}: U_{ijk}\ra H$ are now constant. Thus the equivalence class of the flat 2-connection is completely determined by the constant functions $\sa_{ij}: U_{ij}\ra \Pi_1$ and $h_{ijk}: U_{ijk}\ra H$. On every $U_{ijk}$ these functions satisfy a constraint:
$$
t(h_{ijk})=s(\sa_{ik}) s(\sa_{ji}) s(\sa_{kj}),
$$
which implies that the functions $\sa$ form a Cech 1-cocycle with values in $\Pi_1$:
$$
\sa_{ik}\sa_{ji}\sa_{kj}=1.
$$
Furthermore, $h_{ijk}$ can be expressed in terms of the functions $\sa$ and a constant function $\sb_{ijk}: U_{ijk}\ra \Pi_2$:
$$
h_{ijk}=\sb_{ijk} \tf(\sa_{ik},\sa_{ji}),
$$
where $\tf: \Pi_1\times\Pi_1\ra H$ is some lift of $f:\Pi_1\times \Pi_1\ra \im\ t$ defined by $f(x,x')=s(x)s(x') s(x\cdot x')^{-1}$. The functions $\sb_{ijk}$ form a 2-cochain with values in $\Pi_2$. This cochain is a twisted cocycle, in the sense that it satisfies
\begin{multline}
\alpha(\sa_{kl})(\sb_{ijk})-\sb_{ijl}+\sb_{ikl}-\sb_{jkl}=\\ \tf(\sa_{il},\sa_{ji}) \tf(\sa_{jl}, \sa_{kj}) \left(\tf(\sa_{il},\sa_{ki}) \alpha(s(\sa_{kl}))(\tf(\sa_{ik},\sa_{ji}) \right)^{-1}.
\end{multline}
Here we used additive notation for the group operation on $\Pi_2$, since $\Pi_2$ is an abelian group. Note that the right-hand side takes values in $\Pi_2=\ker\ t$, because 
$f:\Pi_1\times\Pi_1\ra \im\ t$ is a 2-cocycle.

The left-hand side of this equation can be interpreted as a differential $\delta_{\sa}$  in the Cech complex computing the cohomology of a local system on $X$. This local system has $\Pi_2$ as its fiber and is associated to the $\Pi_1$ local system defined by the 1-cocycle $\sa$ via the action of $\Pi_1$ on $\Pi_2$.  The right-hand side also has a nice interpretation, if we recall \cite{Hatcher} that a 1-cocycle $\sa$ with values in $\Pi_1$ can be interpreted as defining a map $\sa: X\ra B\Pi_1$, and that a 2-group $(G,H,t,\alpha)$ defines a class $\beta\in H^3(B\Pi_1,\Pi_2)$. Then one can show that the right-hand side is a 3-cochain representing the class $\sa^*\beta$. 

To summarize, a flat 2-connection is determined by a 1-cocycle $\sa$ with values in $\Pi_1$ and a 2-cochain $\sb$ with values in a $\Pi_2$ local system satisfying the twisted cocycle condition
$$
\delta_{\sa} \sb=\sa^*\beta.
$$

The residual gauge equivalences are described by constant $\Pi_1$-valued functions $f_i$ on each $U_i$ and by constant $\Pi_2$-valued functions $\mu_{ij}$ on each $U_{ij}$. The former transformations act as follows:
$$
\sa_{ij}\mapsto f_j \sa_{ij} f_i^{-1},\quad \sb_{ijk}\mapsto \alpha(f_k)(\sb_{ijk}).
$$
The latter transformations act trivially on $\sa$ and shift the 2-cochain $\sb$ by a coboundary:
$$
\sb_{ijk} \mapsto \sb_{ijk}-\alpha(\sa_{jk}) (\mu_{ij})-\mu_{jk}-\mu_{ki}.
$$

\section{2-gauge theory on a lattice}

The lattice formulation of the $\cG$ gauge theory is most easily defined with respect to a triangulation of the space-time $X$. The simplices need to be oriented so that our formulas have the proper signs. This can be achieved by ordering the vertices. We use the convention that higher vertex labels point towards lower vertex labels since our group elements act on the left. This is demonstrated in figures below. The theory depends on neither the triangulation nor this ordering.

To get a lattice description of the $\cG$ gauge theory we use the description of flat 2-connections in terms of the data $(\Pi_1,\Pi_2,\alpha,\beta)$ obtained in the previous section. All one has to do is to rewrite the cochains $\sa$ and $\sb$ and conditions on them in simplicial terms. Since this is standard, we simply state the results.

A field configuration is an assignment of an element of $A\in \Pi_1$ to each edge (1-simplex) and of an element $B\in \Pi_2$ to each triangle (2-simplex). These correspond to the cochains $\sa$ and $\sb$ of the previous section. They are subject to the following constraints. Given a triangle with edge variables $A_0,A_1,A_2$, labeled in the ordinary way according to the induced ordering on the vertices we have a flatness constraint
\begin{equation}\label{flatness}
A_0A_1^{-1}A_2 = 1,
\end{equation}
where the labeling and orientation is demonstrated in the figure below.

\begin{center}
\begin{tikzpicture}
[scale=1, vertices/.style={draw, fill=black, circle, inner sep=0.5pt}]
\node[vertices, label=right:{$1$}] (1) at (0.8,0.8) {};
\node[vertices, label=left:{$0$}] (0) at (-1,0.5) {};
\node[vertices, label=above:{$2$}] (2) at (-0.1,1.5) {};
\draw[->-=0.55] (2)--(1) node at (0.6,1.4) {$A_0$};
\draw[->-=0.5] (2)--(0) node at (-0.75,1.35) {$A_1$};
\draw [->-=0.5] (1)--(0) node at (0,0.3) {$A_2$};
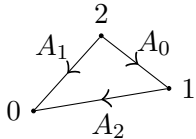
\node at (0,-1) {Labeling for edge variables on 2-simplices.};
\end{tikzpicture}
\end{center}
This ensures that the holonomy of $A$ around any closed curve only depends on the homotopy class of that curve.
Note that  group elements are thought of as acting from the left.

There is also a constraint for every tetrahedron (3-simplex). For a tetrahedron with face variables $B_0,B_1,B_2,B_3\in \Pi_2$ labeled in the ordinary way and edge variables $A_0,A_1,A_2 \in \Pi_1$ labeled as shown in the picture below, one requires:

\begin{equation}\label{2-flatness}
\alpha(A_0)(B_0)-B_1+B_2-B_3 = \beta(A_0,A_1,A_2).
\end{equation}

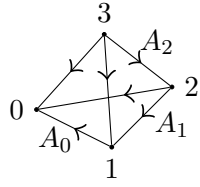
\begin{center}
\begin{tikzpicture}
[scale=1, vertices/.style={draw, fill=black, circle, inner sep=0.5pt}]
\node[vertices, label=below:{$1$}] (1) at (0,0) {};
\node[vertices, label=right:{$2$}] (2) at (0.8,0.8) {};
\node[vertices, label=left:{$0$}] (0) at (-1,0.5) {};
\node[vertices, label=above:{$3$}] (3) at (-0.1,1.5) {};
\draw[->-=0.5] (2)--(1) node at (0.8,0.3) {$A_1$};
\draw[->-=0.5] (1)--(0) node at (-0.75,0.1) {$A_0$};
\draw[->-=0.4] (3)--(1);
\draw[->-=0.55] (3)--(2) node at (0.6,1.4) {$A_2$};
\draw[->-=0.5] (3)--(0);
\draw [->-=0.35] (2)--(0);
\node at (0,-1.5) {Labeling for edge variables on 3-simplices.};
\end{tikzpicture}
\end{center}
Using the differential $d_A$ on twisted cochains this constraint can be written as
\begin{equation}\label{2-flatnessprime}
d_A B=\beta(A). 
\end{equation}

There are two kinds of gauge transformations for this theory.  First, there are ordinary ( 0-form) gauge transformations depending on a 0-cochain $f$ with values in $\Pi_1$. Such a transformation acts on $A$ as follows:
\begin{equation}\label{Af}
A\mapsto A^f,\quad A_\gamma\mapsto A_\gamma^f = f_0A_\gamma f_1^{-1},
\end{equation}
where $A_\gamma$ is the value of $A$ on an oriented 1-simplex $\gamma$, and $f_0$ and $f_1$ are the values of $f$ on the two endpoints of $\gamma$.
These transformations also affect the $B$ field:
\begin{equation}\label{Bf}
B_\Sigma\mapsto B_\Sigma^f = \alpha(f_0)(B_\Sigma) + \zeta_\Sigma(A,f)
\end{equation}
on each 2-simplex $\Sigma$. Here $\zeta$ is a $\Pi_2$-valued 2-cochain satisfying
\begin{equation}\label{zeta}
d_{A^f}\zeta(A,f) = \beta(A^f) - \alpha(f)(\beta(A)).
\end{equation}
In what follows we adopt a simplified notation for the action of a 0-cochain $f$ with values in $\Pi_1$ on a $p$-cochain $\nu$ with values in a local system with fiber $\Pi_2$:   instead of
$\alpha(f)(\nu)$ we write $f\cdot \nu$. Thus the definition of $\zeta$ takes the form
\begin{equation}\label{zetaprime}
d_{A^f}\zeta(A,f) = \beta(A^f) - f\cdot \beta(A),
\end{equation}
and the action of $f$ on $B$ takes the form
\begin{equation}\label{Bfprime}
B\mapsto B^f=f\cdot B+\zeta(A,f).
\end{equation}
The inhomogeneous term in the transformation (\ref{Bfprime}) is necessary to preserve the twisted flatness condition (\ref{2-flatnessprime}). Indeed:
\begin{equation}\label{cocyclecheck}
d_{A^f} B^f = d_{A^f}(f\cdot B + \zeta(A,f)) = d_{A^f}(f\cdot B) + \beta(A^f) - f\cdot \beta(A).
\end{equation}
A simple calculation shows
$$
d_{A^f}( f\cdot B) = f\cdot d_{A} B = f \cdot \beta(A),
$$
hence the pair $(B^f,A^f)$ satisfies the condition (\ref{2-flatnessprime}).

Note that the equation (\ref{zetaprime}) for $\zeta$ always has a solution, since the cohomology class of $\beta(A)$ does not change when one replaces $A$ by a gauge-equivalent one (this follows from the fact that $\beta$ is closed). We can even choose $\zeta$ to satisfy the normalization condition
\begin{equation}\label{zetanorm}
\zeta(1^g,f) = 0,
\end{equation}
for all constant $f \in \Pi_1$ and gauge-trivial $A = 1^g$. This normalization is possible since if $f$ is constant, the cochain $\beta$ can be normalized so that $\beta(1^{gf}) = f\cdot \beta(1^g)$.

The second class of gauge transformations are 1-form gauge transformations depending on a 1-cochain $a$ with values in $\Pi_2$. These transformations do not affect the 1-cocycle $A$, while   $B$ transforms as follows:
\begin{equation}\label{Ba}
B_\Sigma\mapsto B_\Sigma^a = B_\Sigma + \alpha(A_2) (a_0) - a_1 + a_2,
\end{equation}
for every 2-simplex $\Sigma$ whose  boundary $\partial\Sigma$ is assigned $a_0, a_1, a_2$. In other words, $B\mapsto B^a=B+d_A a$.  These transformations preserve the twisted flatness constraint for $B$ since $d_A^2 = 0$.

The 1-form gauge transformations have 2-gauge transformations parametrized by an assignment of an element of $\Pi_2$ to each vertex under which the 1-form gauge parameters $a$ transform as
\begin{equation}\label{ag}
a_\gamma^g = a_\gamma + g_1-\alpha(A_\gamma)(g_0),
\end{equation}
where $\partial\gamma$ is assigned $g_0, g_1$. One can check that two 1-form gauge transformations related by such a 2-gauge transformation act identically on the fields.

1-form and 0-form gauge transformations do not commute in general. Let $a$ be a 1-cochain with values in the $\Pi_2$ local system associated to a $\Pi_1$ local system, and $f$ be a 0-cochain with values in $\Pi_1$. Let us denote by $(a,f)$ as a 1-form transformation with a parameter $a$ followed by a 0-form transformation with a parameter $f$. Then
\begin{equation}\label{commtransf}
(a,f)=(0,f)\circ (a,1)=(f\cdot a,1)\circ (0,f).
\end{equation}
For future use, we record the transformations of $B$ and $A$ under a general transformation $(a,f)$: 
\begin{equation}\label{Baf}
B^{a,f} = f\cdot (B + d_{A} a) + \zeta(A,f),
\end{equation}
and 
\begin{equation}\label{Aaf}
A^{a,f} = A^f.
\end{equation}

When composing two gauge transformations, where the first one is $(a_1,f_1)$, it is often convenient to write the second one as $(f_1\cdot a_2, f_2)$ rather than $(a_2,f_2)$. We will use this convention below.

The reader might have noticed the similarity between $\beta$ and $\zeta$ on one hand, and the Chern-Simons form and its descendant on the other hand. This is not a coincidence, since the Chern-Simons form for a compact Lie group $G$ can be regarded as a 3-cocycle on the classifying space of flat $G$-connections \cite{CheegerSimons}. In superstring theory and supergravity, the Chern-Simons form modifies the definition of the gauge-invariant field-strength of a 2-form gauge field $B$, while the descendant of the Chern-Simons form describes how the $B$-field transforms under ordinary gauge symmetries \cite{Polchinski}. This is clearly analogous to how $\beta$ modifies the flatness constraint on $B$ (eq. (\ref{2-flatnessprime})) and how $\zeta$ enters the transformation law for $B$ (eq. (\ref{Bf})). 

It is well-known that the descent procedure can be continued, so that from the 2-cochain $\zeta(A,f)$ one gets a 1-cochain $\kappa(A,f,g)$, etc. Below we will need the definition of $\kappa$ only for $A=0$, so let us describe how it is constructed. By definition,
$$
\beta(A^g)-g\cdot \beta(A)=d_{A^g} \zeta(A,g),
$$
where $A$ is an arbitrary 1-cochain with values in $\Pi_1$ and $g$ is an arbitrary 0-cochain with values in $\Pi_1$. 
Let us set $A=1^f$, where $f$ is an arbitrary 0-cochain with values in $\Pi_1$  (i.e. we let $A$ be cohomologous to the trivial 1-cochain). Then we get
$$
\beta(1^{gf})-g\cdot \beta(1^f)=d_{1^{gf}} \zeta(1^f,g)=gf \cdot d\left((gf)^{-1} \cdot \zeta(1^f,g)\right).
$$
This equation implies that the 2-cochain 
$$
(gf)^{-1} \cdot \zeta(1,gf)-f^{-1}\cdot \zeta(1,f)-(gf)^{-1}\cdot \zeta(1^f,g)
$$
is closed for any 0-cochains $f$ and $g$. Contractibility of $EG$ implies that this cochain is exact, i.e. there exists a 1-cochain $\kappa(g,f)$ with values in $\Pi_2$ such that
\begin{equation}\label{kappa}
(gf) d\kappa(g,f)=g\cdot\kappa(1,f)+\zeta(1^f,g)-\zeta(1,gf).
\end{equation}
This 1-cochain $\kappa$ is the second descendant of $\beta$.

The partition function is defined as the weighted sum over all allowed field configurations, divided by the order of the group of gauge transformations and multiplied by the order of the group of 2-gauge transformations. The weight must be gauge-invariant and topologically-invariant, in the sense that the partition function must be invariant under a subdivision of the triangulation. The most general weight function satisfying these conditions will be described in section \ref{top_actions}.

\section{Loop and surface observables}

The above construction of the 2-group TQFT works in arbitrary space-time dimension (and gives something different from the Dijkgraaf-Witten theory in dimension $3$ or higher). We now discuss observables in this TQFT focusing on the 4d case. 

The 2-group TQFT in 4d describes the phase with both electric gauge group $\Pi_1$ and magnetic gauge group $\Pi_2$ and therefore  admits two types of loop observables and two types of surface observables.
Let us begin with loop observables. There are ordinary Wilson loops for the lattice gauge field $A$; they are labeled by representations of $\Pi_1$. There are also disorder loop operators which correspond to 't Hooft loops in the microscopic gauge theory. In the lattice formulation, one chooses a closed path $\gamma$ on the dual cell complex and modifies the twisted cocycle condition (\ref{2-flatness}) on every 3-simplex $T_l$ dual to an edge of $l\in\gamma$ as follows:
$$
\alpha(A_0)(B_0)-B_1+B_2-B_3 = \beta(A_0,A_1,A_2)+H_l.
$$ 
Here $H_l$ is an element of $\Pi_2$. One can think of the elements $H_l$ as defining a dual 1-cochain with values in a $\Pi_2$ local system on the loop $\gamma$. The above equation implies that this cochain is a cocycle, so the elements $H_l$ for different $l$ all lie in the same orbit of the $\Pi_1$ action on $\Pi_2$.
Therefore 't Hooft loops are labeled by orbits of the  $\Pi_1$-action in $\Pi_2$. 

There are also order and disorder surface observables.  Disorder surface observables are defined by the condition that the $\Pi_1$ gauge field has a fixed holonomy along a loop linking a homologically trivial surface $\Sigma$. In the lattice formulation, $\Sigma$ should be thought of as composed of 2-cells of the dual cell complex, and the insertion of the disorder operator supported on $\Sigma$ amounts to deforming the 1-cocycle condition on $A$ for every 2-simplex dual to the 2-cells of $\Sigma$. Such surface observables are labeled by conjugacy classes in $\Pi_1$.

There are also Wilson surface observables are labeled by elements of $\hPi_2=\Hom(\Pi_2,U(1))$ which satisfy two conditions. First of all, $\eta\in \hPi_2$ must be invariant under the action of $\Pi_1$. Second, $\eta\circ\beta\in H^3(\Pi_1,U(1))$ must vanish. 

A quick way to see how these two conditions on $\eta$  arise is as follows. The Wilson surface measures the flux of $B$ though $\Sigma$. Thus it must involve a product over all 2-simplices making up $\Sigma$, with each simplex contributing a phase $\eta(\pm B)$, where the sign is determined by the mutual orientation of $\Sigma$ and the 2-simplex in question. Invariance with respect to gauge transformations requires $\eta$ to $\Pi_1$-invariant, giving the first condition on $\eta$. To obtain the second condition, note that the Wilson surface observable should evaluate to $1$ if $\Sigma$ is the boundary of a 3-simplex and no 't Hooft loop intersects this 3-simplex. Applying $\eta$ to the twisted cocycle condition (\ref{2-flatness}) we see that this is true if $\eta\circ\beta(A_0,A_1,A_2)=0$ for all $A_0,A_1,A_2\in\Pi_1$. More generally, suppose the 3-cochain $\eta\circ\beta(A_0,A_1,A_2)$ is not zero but is a coboundary of a 2-cochain on $\Pi_1$ with values in $U(1)$:
$$
\eta\circ\beta(A_0,A_1,A_2)=\gamma(A_0,A_1A_2)\gamma(A_0 A_1, A_2)^{-1}\gamma(A_1,A_2)\gamma(A_0,A_1)^{-1}.
$$
Then one can modify the definition of the Wilson surface observable by multiplying the weight assigned to a 2-simplex by a factor $\gamma(A_0,A_1)$. The modified observable is trivial on an elementary 3-simplex, as required. On the other hand, if $\eta\circ\beta$ is not cohomologous to zero, no local modification of the weight can solve the problem. 

Apart from the surface observables measuring the flux of $B$, one can also construct  surface observables which are sensitive only to the 1-cocycle $A$. Namely, if we restrict $A$ to a closed surface $\Sigma$, it defines a principal $G$-bundle on $\Sigma$ with structure group $\Pi_1$, and any class in $H^2(B\Pi_1,U(1))$ gives a surface observable. Such observables are trivial if $\Sigma$ is simply-connected, but are nontrivial in general.

The most general observable supported on a surface $\Sigma$ involves both the $B$ fluxes and the $A$-holonomies along a marked 1-skeleton of $\Sigma$. The $\Pi_1$ invariance of such operators may rely on the transformation of both $A$ and $B$ variables, so they are not necessarily products of Wilson lines and Wilson surfaces.

In three dimensions the analysis is very similar, so we just present the results. There are local operators (i.e. operators supported at points) labeled by orbits of $\Pi_1$ action on $\Pi_2$. These correspond to 't Hooft point operators in the microscopic gauge theory. There are two kinds of loop observables: Wilson loops labeled by representations of $\Pi_1$ and vortex loops labeled by conjugacy classes in $\Pi_1$. Finally there are Wilson surfaces defined in the same way as in four dimensions. That is, they are labeled by a $\Pi_1$-invariant element $\eta\in \hPi_2$ which annihilates the class $\beta$, together with a class in $H^2(B\Pi_1,U(1))$.

\section{Topological actions}\label{top_actions}

As described in the appendix, the configuration data $(A,B)$ are equivalent to a simplicial map $\mathcal{B}$ from the spacetime $X$ into the classifying space $B\cG$ of the 2-group. Gauge equivalence classes are homotopy classes of these maps. This is analogous to the fact that a flat $G$-connection can be viewed as a map to the classifying space $BG$. We therefore define an action functional for a $d$-dimensional theory by picking a class $\mathcal{L} \in H^d(B\cG,\RZ)$ and setting
\begin{equation}\label{action}
S(\mathcal{B}) = 2\pi i \int_X \mathcal{B}^*\mathcal{L}.
\end{equation}
This action is gauge invariant and manifestly topological.

To calculate the cohomology group $H^d(B\cG,\RZ)$ we need a good understanding of the classifying space $B\cG$. This space can be taken to be any space with $\cG$ as its homotopy type. This implies $\pi_1(B\cG) = \Pi_1$, $\pi_2(B\cG)=\Pi_2$, and all higher homotopy groups vanish. It also means that the action of $\pi_1(B\cG)$ on $\pi_2(B\cG)$ is given by $\alpha$ as well as a more complicated condition involving the Postnikov invariant $\beta$. In the appendix we describe a cell structure for this space.

It turns out $B\cG$ is a fibration over $B\Pi_1$ with fiber $B^2\Pi_2$:

\begin{center}
\begin{tikzpicture}
\node (BG) {$B\cG$};
\node (Bpi1) [below of=BG, node distance=1.1cm] {$B\Pi_1$};
\node(X) [left of=BG, node distance=1.5cm] {$B^2\Pi_2$};
\draw[->] (BG) to node {} (Bpi1);
\draw[->] (X) to node {} (BG);
\end{tikzpicture}
\end{center}
Here $B^2\Pi_2$ is a space defined up to homotopy by the condition that $\pi_2(B^2\Pi_2)=\Pi_2$ while all other homotopy groups vanish.
This fibration is classified by the Postnikov class $\beta \in H^3(B\Pi_1,\Pi_2)$. We know the cohomology of the base and fibers so we can use the Serre spectral sequence to get a reasonable handle on the cohomology of $B\cG$. Details of this spectral sequence are in the appendix.

The Serre spectral sequence tells us that for 2-dimensional theories, $\mathcal{L}$ is a sum of two terms. The first term is an element of $H^2(B^2\Pi_2,\RZ)$. This cohomology group is isomorphic to the group $\hat{\Pi}_2$ of homomorphisms $\Pi_2 \to \RZ$. A character $\eta \in \hat{\Pi}_2$ gives a class on $B\cG$ if and only if it is invariant under the action of $\Pi_1$ and $\eta(\beta) \in H^3(B\Pi_1,\RZ)$ is zero as a cohomology class. An interpretation of these conditions was discussed in the previous section.

The second term can be any element $\omega \in H^2(B\Pi_1,\RZ)$.

The most general 2d action is therefore
$$
S(\mathcal{B}) = 2\pi i \int_X \eta(B) + 2\pi i\int_X\mathcal{A}^*\omega,
$$
where here $B$ is the $\Pi_2$-valued 2-form field and $\mathcal{A}:X\ra B\Pi_1$ is the composition of the classifying map $\mathcal{B}:X \to B\cG$ with the fibration map $B\cG \to B\Pi_1$. Equivalently this is the map $X \to B\Pi_1$ determined by the $A$ variables. Note that if $X = S^2$, then since $\pi_2(B\Pi_1)=0$, the pullback $\mathcal{A}^*\omega$ always integrates to zero.

In three dimensions, we also find that $\mathcal{L}$ is a sum of two terms. The first term is an element $\omega \in H^3(B\Pi_1,\RZ)$. The action depends on this element only up to the addition of a term of the form $\eta(\beta)$, where $\eta$ is a $\Pi_1$-invariant character of $\Pi_2$. 

The second term is an element $\lambda\in H^1(B\Pi_1,\hat\Pi_2)$. As a cochain on $B\cG$, $\lambda$ is closed only if $\langle\lambda,\cup\beta\rangle \in H^4(B\Pi_1,\RZ)$ vanishes as a cohomology class.

The most general 3d action is therefore
$$
S(\mathcal{B}) = 2\pi i \int_X \langle \mathcal{A}^*\lambda, \cup B \rangle + 2\pi i \int_X \mathcal{A}^*\omega.
$$
Note that on a simply-connected space $\omega$ does not contribute since $\mathcal{A}^*\omega$ is exact. On the other hand, athough $\mathcal{A}^*\lambda$ is exact, $B$ is not closed, so their contraction is not necessarily exact.

In four dimensions the action is a sum of three terms. The first term depends on an element of $H^4(B^2\Pi_2,\RZ)$. This group is isomorphic to the group of quadratic functions $q:\Pi_2 \to \RZ$ as discussed in \cite{KT1}. In order for this cocycle to extend to $B\cG$, it needs to be invariant under the action of $\Pi_1$. This is equivalent to the quadratic function $q$ being invariant. In order for its extension to be closed, $\langle \beta, - \rangle_q$ needs to vanish as an element of $H^3(B\Pi_1,\hat\Pi_2)$, where the bracket denotes the associated bilinear form for $q$.

The second term depends on an element $\lambda\in H^2(B\Pi_1,\hat{\Pi}_2)$. This gives a cocycle on $B\cG$ when contracted with the $B$-field only if $\langle\lambda,\cup\beta\rangle$ is zero in $H^5(B\Pi_1,\RZ)$.

The third term is an integral of a pull-back of $\omega \in H^4(B\Pi_1,\RZ)$. The action only depends on $\omega$ up to the addition of something of the form $\langle\gamma,\cup\beta\rangle$ for some $\gamma \in H^1(B\Pi_1,\hat\Pi_2)$.

The most general 4d action is therefore
$$
S(\mathcal{B}) = 2\pi i \int_X q_*(\mathfrak{P}B)+2\pi i \int_X \langle \mathcal{A}^*\lambda , \cup B\rangle +2\pi i \int_X \mathcal{A}^*\omega,
$$
where $\mathfrak{P}B$ denotes the Pontryagin square of the $B$-field, and $q_*$ is the map from the universal quadratic group $\Gamma(\Pi_2)$ to $\RZ$ corresponding to $q$. This term is discussed in detail in \cite{KT1}. Note that because of the Postnikov class $\beta$, it is only closed after applying $q_*$. For simply-connected $X$, as in the three-dimensional case, $\omega$ does not contribute, but $\lambda$ does.

In the case of Dijkgraaf-Witten theory in dimension $d$, the topological action depends on a class in $H^d(BG,U(1))$, and one can give an explicit description of the corresponding cocycle as a function of $d$ variables living in $G$. From the mathematical viewpoint, this explicit description arises from the standard complex computing the cohomology of $BG$, while from the physical viewpoint the cocycle is the weight attached to a $d$-simplex \cite{DW}. Similarly, one can give an explicit description of a class in $H^d(B\cG,U(1))$ as a function of several variables, some of them living in $\Pi_1$ and some living in $\Pi_2$, satisfying a certain condition. This condition can be understood mathematically as a cocycle condition in a standard complex computing the cohomology of $B\cG$. However, even in low degrees the formulas are quite unwieldy. For this reason we only state them for $d=2$. In this case the cocycle is a function $\cL$ on $\Pi_1\times\Pi_1\times\Pi_2$ with values in $\RR/\ZZ$. The cocycle condition is
$$
\cL(A_0,A_1A_2; B_2)-\cL(A_0 A_1, A_2; B_1)-\cL(A_0, A_1; B_3)+\cL(A_1, A_2; B_0)=0,
$$
where $B_0$ is not independent but is expressed through the variables $A_0,A_1,A_2, B_1, B_2, B_3$ by means of (\ref{2-flatness}). 2-cocycles for $\Pi_1$ can be identified with 2-cocycles for $B\cG$ which do not depend on the $\Pi_2$ variable. 

A 3-cocycle for a 2-group $B\cG$ depends on three variables in $\Pi_1$ and three variables in $\Pi_2$. In general, a $d$-cocycle depends on $d$ variables in $\Pi_1$ and $d(d-1)/2$ variables in $\Pi_2$. They can be thought of as labeling edge and triangles containing a given vertex of a $d$-simplex.

\section{Duality}

Since the simplicial $B$-field takes values in an abelian group $\Pi_2$, one could try to dualize it. In dimension $d$ the dual variable should live on $(d-3)$-cells of the dual cell complex and take values in the Pontryagin-dual group $\hPi_2=\Hom(\Pi_2,U(1))$. But in general a nontrivial topological action for $B$ obstructs the dualization procedure. An important special case where the dualization can be performed is when the action is either independent of $B$, or depends on it linearly. Let us perform the dualization procedure in various dimensions. 

For $d=2$ the 2-group TQFT is essentially equivalent to the Dijkgraaf-Witten theory with gauge group $\Pi_1$. First of all, the constraint (\ref{2-flatness}) is not needed in this case. Second, the class $\eta$ in the 2d action must vanish, because otherwise the partition function on $S^2$ vanishes after one sums over the $B$-fields, which contradicts the axioms of TQFT \cite{WittenCS}. But then summation over $B$ only produces an inessential numerical factor, and the theory is clearly equivalent to the Dikgraaf-Witten theory for $\Pi_1$ with an action given by $\omega\in H^2(B\Pi_1,\RR/\ZZ)$. 

For $d=3$ the action is always linear in $B$, as explained in the previous section. Therefore the dualization is always possible. We impose the constraint (\ref{2-flatness}) by means of a Lagrange multiplier field $C$ which lives on dual 0-cells and takes values in $\hPi_2$. Thus we add to the action a term
$$
2\pi i \int_X \langle C, \delta_A B-\beta(A) \rangle
$$
and treat $B$ as an unconstrained 2-cochain with values in $\Pi_2$.
For simplicity, let us first assume that $\Pi_1$ acts trivially on $\Pi_2$. Then the differential $\delta_A$ becomes the usual Cech differential $\delta$, and summing over $B$ produces a constraint 
$$\delta C= -\cA^*\lambda,$$ where $\lambda\in H^1(B\Pi_1,\hPi_2)$. 

Since we assumed for now that $\alpha$ is trivial, the group $H^1(B\Pi_1,\hPi_2)$ is merely the group of homomorphisms from $\Pi_1$ to $\hPi_2$. Thus the constraint on $C$ reads explicitly
$$
C_1-C_0+\lambda(A_l)=0,
$$
where $l$ is a 1-cell of the  of the dual cell complex with source and target $0$ and $1$.\footnote{Here we neglected the fact that the field $A$ was originally only defined on the 1-cells of the original triangulation. In more detail, one needs first to pass from the simplicial 1-cochain $A$ to a $\Pi_1$ local system on $X$ and then restrict to the 1-cells of the dual cell complex. The first step involves a certain arbitrariness which does not change the conclusions.} This constraint is gauge-invariant provided we assign to $C$ a nontrivial transformation law under $\Pi_1$ gauge transformations:
$$
C\mapsto C+\lambda(f),
$$
where $f$ is a 0-cochain with values in $\Pi_1$ parameterizing a gauge transformation. Thus the dual theory is a topological sigma-model with target $\hPi_2$ coupled to a topological 3d gauge theory with gauge group $\Pi_1$. The group $\Pi_1$ acts on $\hPi_2$ via the homomorphism $\lambda$. The action of the gauge theory is 
$$
S=2\pi i \int_X \cA^*\omega+2\pi i\int_X \langle C, \cA^*\beta \rangle.
$$
Note that while the first term is the Dijkgraaf-Witten action of the 3d gauge theory, the second term is of a different nature. 

The case of nontrivial $\alpha$ is not very different. The constraint on $C$ now reads 
$$
\delta_A C+\lambda(A)=0.
$$
The transformation law for $C$ is now more complicated:
$$
C\mapsto \hat\alpha(f)(C)+\lambda(f),
$$
where $\hat\alpha$ denotes the action of $\Pi_1$ on $\hPi_2$ dual to the action of $\Pi_1$ on $\Pi_2$. The dual theory is again a gauged topological sigma-model with the same action as above.

For $d=4$ we need to assume that the first term in the topological action vanishes (i.e. $q=0$). Then the action is again linear in $B$, and $B$ can be dualized to a 1-cochain $C$ with values in $\hPi_2$. Performing summation over $B$ we find a constraint on $C$ which reads
$$
\delta_A C=\lambda(A),
$$
where $\lambda$ is a cocycle representing a class in $H^2(\Pi_1,\hPi_2)$. We also have the usual constraint which says that $A$ is a 1-cocycle with values in $\Pi_1$. One can show that these two constraints can be interpreted as the flatness constraint for a gauge field $(A,C)$ taking values in an extension of $\Pi_1$ by $\hPi_2$ determined by $\alpha$ and $\lambda$. Thus the dual theory is a topological gauge theory with this extension as the gauge group. The action is necessarily of the DW type and has the form
$$
S=2\pi i \int_X \cA^* \omega+2\pi i \int_X \langle C, \cup\cA^*\beta\rangle. 
$$
Note that even if the original 2-group TQFT had a trivial action (i.e. $\lambda$ and $\omega$ vanish), the dual theory has a nontrivial action which is determined by $\beta$. On the other hand, the class $\lambda$ which parameterized the action of the original theory enters the dual theory only through the structure of the gauge group.

\section{Phases protected by higher symmetry}

\subsection{SPT phases and TQFT}

Recently, the Dijkgraaf-Witten TQFT was used in a novel way, as a tool to classify symmetry-protected (SPT) phases without long-range entanglement \cite{SPT, LevinGu, SPT2}. These are phases of matter which have three properties: (1) they are gapped; (2) they have a global symmetry group $G$ which acts ultralocally; (3) the TQFT describing the low-energy limit is trivial. Here by an ultralocal action of a global symmetry we mean that there is a lattice realization of the phase where the symmetry transformation only mixes degrees of freedom living on a given vertex (0-cell).  

It was proposed in \cite{SPT} that SPT phases in space-time dimension $d$ with a finite internal \footnote{We will focus on the case of finite internal symmetry, but one can generalize it to the case when $G$ is a compact Lie group which might involve time-reversal.}  symmetry group $G$ are classified by elements of $H^d(BG,U(1))$. Let us provide an interpretation of this classification scheme in TQFT terms (see also \cite{SPT2} for a very similar discussion). We use the fact that a lattice system with an ultralocal internal symmetry can be canonically coupled to a flat gauge field with structure group $G$. Indeed, locally any flat $G$-connection is a pure gauge, i.e. a gauge transformation of the trivial connection.  This gauge transformation is defined up to a constant symmetry transformation in $G$. Since the system is local, the lattice action is a sum over all vertices
$$
\sum_v S_v(\phi),
$$
where each term $S_v$ depends only on the degrees of freedom in the immediate neighborhood of the vertex. Since the symmetry acts ultralocally, each $S_v$ is separately invariant under constant symmetry transformations. We define the action of the system coupled to a flat background gauge field as a sum 
$$
\sum_v S_v(\phi^g),
$$ 
where $\phi^g$ is a transformation of the field configuration by a local gauge transformation $g$ describing the flat connection. Since different choices of $g$ differ by constant symmetry transformations, this expression is well-defined and gauge-covariant.  Now we can integrate out the matter fields and obtain an effective action for the flat gauge field. It is necessarily topological and therefore must arise from a class in $H^d(BG,U(1))$.

It is natural to ask whether more general 2-group TQFTs we have studied here and in \cite{GK, KT1} are related to new phases of matter not covered by the group cohomology classification. The role of the symmetry group $G$ is taken by a 2-group $\cG$. In this more general setting, a global symmetry transformation is parameterized by an element of $\Pi_1$ and a flat gauge field with gauge group $\Pi_2$. The action of $\cG$ is assumed to be ultralocal, in the sense that the system can be canonically coupled to a flat 2-connection $(A,B)$ with structure group $\cG$.  If the system is gapped and the ground state has short-range entanglement, one can integrate out the matter fields and obtain a topological action for $(A,B)$ which is described by an element of $H^d(B\cG,U(1))$. 

In the case when $\Pi_2$ is trivial, this reduces to the group cohomology classification of SPT phases. The opposite extreme is when $\Pi_1$ is trivial. In this case $\cG$ is completely determined by an abelian group $\Pi_2$, and phases protected by such a 2-symmetry are labeled by $H^d(B^2\Pi_2,U(1))$. The first nontrivial case is $d=4$, where $\Pi_2$-protected phases are classified by quadratic functions on $\Pi_2$ with values in $U(1)$. The same data classify pre-modular braided tensor categories whose simple objects are labeled by elements of $\Pi_2$. Since pre-modular categories are also used in the construction of Walker-Wang TQFTs \cite{WW}, it seems likely that 2-group TQFTs with trivial $\Pi_1$ are a special case of Walker-Wang models. Indeed, it has been conjectured in \cite{WW} (see also \cite{KBS}) that in the continuum limit Walker-Wang models based on $\ZZ_n$  are described by a $BF$ action deformed by a term $B\wedge B$. This agrees with the continuum description of the 2-group TQFT discussed in \cite{GK, KT1}. 

A simple example of a system with an ultralocal 2-symmetry is given by a Yang-Mills theory with gauge group $G$ where all matter fields transform trivially under the center of $G$. Then the symmetry 2-group has $\Pi_1=0$ and $\Pi_2=Z(G)$. The system can be ``minimally coupled'' to a flat B-field with values in $Z(G)$. Essentially, this means that one performs the path-integral over Yang-Mills gauge fields with structure group $G/Z(G)$ and a fixed 't Hooft flux described by $B$. The resulting function of $B$ must be an integral over $X$ of a pull-back of a class in $H^d(B^2 Z(G),U(1))$. In the case $d=4$, one can interpret the elements of this group as labeling discrete theta-angles in the underlying Yang-Mills theory \cite{AST,KT1}.

Even more generally, if one dealing with a gapped phase in space-dimension $d$, symmetry transformations may live on cells of all dimensions up to $d$. This situation is most natural when the ``matter fields'' involve  gauge fields of all form degrees. The symmetry structure in this case is described by a $d$-group, i.e. by a $d$-category with a single object and invertible 1-morphisms, 2-morphisms, etc. SPT phases in $d$ dimensions with symmetry $d$-group $\cG$ should be classified by degree-$d$ cohomology of the classifying space of $\cG$. This classifying space has homotopy groups which may be non-vanishing in degrees up to $d$.

\subsection{Boundaries of SPT phases}

In the case of an SPT phase protected by a finite symmetry group $G$, the nontriviality of the corresponding class in $H^d(BG,U(1))$ has interesting physical consequences for the boundary behavior of the phase \cite{LevinGu}: the boundary cannot be gapped without either breaking symmetry or introducing long-range entanglement. The same appears to be true in the case of phases protected by a $d$-group symmetry. Namely, the boundary cannot be gapped by any perturbation which preserves $\cG$ as an ultralocal symmetry and does not create long-range entanglement. Indeed, if such a perturbation existed, one could couple the system to a flat $d$-connection and integrate out the matter fields even in the presence of a nonempty boundary. This should produce a topological action for the flat $d$-connection which is is gauge-invariant on a $d$-manifold with a boundary. Such an action should have the form
$$
S=2\pi i \int_X \cA^*\omega-2\pi i \int_{\partial X} \cA^*\psi
$$
for some $\psi\in H^{d-1}(B\cG,\RR/\ZZ)$. But this action is gauge-invariant only if $\omega=\delta\psi$, which contradicts the nontriviality of $\omega\in H^d(B\cG,\RR/\ZZ)$. Thus if $\cG$ is preserved on the boundary, the boundary can be gapped only at the expense of creating a topological order on the boundary. Moreover, neither the bulk action, nor the boundary TQFT action are separately gauge-invariant in this case (there is an ``anomaly-inflow'' from the bulk to the boundary). Thus $\cG$ is realized anomalously on the boundary, and this anomaly is measured by $\omega$. 

Alternatively, if we insist on having no topological order on the boundary, $\cG$ must be explicitly or spontaneously broken there, and we can describe possible patterns of symmetry breaking. Namely, $\cG$ must be broken down to a $d$-subgroup $\bbH$ such that $\omega$ becomes exact when restricted to $\bbH$, $\omega\vert_{\bbH}=\delta\psi$. Only then is it possible to write a gauge-invariant effective action in the presence of a boundary.

\subsection{Lattice realization of an SPT phase}

As in \cite{SPT}, we can give descriptions of ungauged ground states that realize this phase in the ``group cohomology" basis. Suppose that $\cG$ is a 2-group described by a quadruple $(\Pi_1, \Pi_2,\alpha,\beta)$, as above.\footnote{The case of a general $d$-group is algebraically more involved.} Let $M$ denote the spatial manifold. We will compute the ground state on $M$ by performing the path integral in the TQFT over a spacetime with boundary $M$. For example, if $M$ is a sphere, we take our spacetime $B$ to be a ball. For these theories, spacetime does not need to be smooth, so we can always take spacetime to be $CM$, the cone over $M$.

A map from $CM$ to $B\cG$ is the same thing as a map from $M$ to $B\cG$ along with a nullhomotopy. In other words it is a gauge transformation for a $\cG$ gauge field living just on the spatial slice $M$. This gives us a way of describing the map from the cone purely in terms of configurations on $M$ and for $\cG$ an ordinary group, and in this way we reproduce the description of ground states in \cite{SPT}.

Concretely, a configuration of the ``matter field" $\Phi$ will be an assignment of an element of $\Pi_1$ to every vertex, an element of $\Pi_2$ to every edge, $\Pi_3$ to every face, and so on, for which a single $d$-simplex is pictured below with 2-group labeling, with $\phi_i \in \Pi_1$, $a_i \in \Pi_2$.

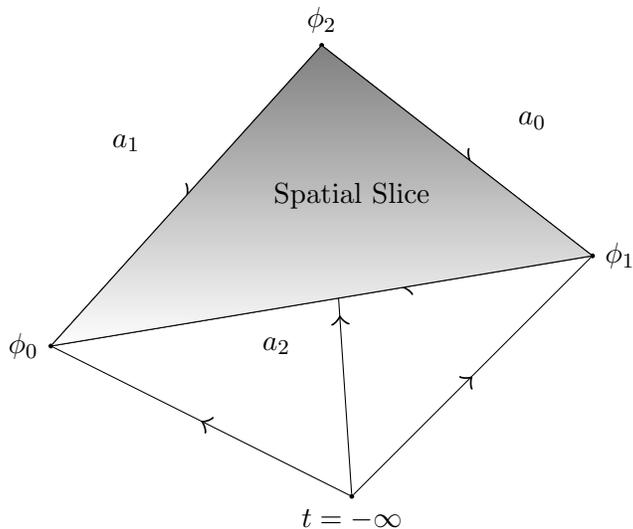
\begin{center}
\begin{tikzpicture}
[scale=1, vertices/.style={draw, fill=black, circle, inner sep=0.5pt}]
\node[vertices, label=below:{$t = -\infty$}] (1) at (0,0) {};
\node[vertices, label=right:{$\phi_1$}] (2) at (3.2,3.2) {};
\node[vertices, label=left:{$\phi_0$}] (0) at (-4,2) {};
\node[vertices, label=above:{$\phi_2$}] (3) at (-0.4,6) {};
\node at (-1,2) {$a_2$};
\draw[->-=0.5] (1)--(2) ;
\draw[->-=0.5] (1)--(0) ;
\draw[->-=0.4] (1)--(3) node at (-3,4.7) {$a_1$};
\draw[->-=0.55] (3)--(2) node at (2.4,5) {$a_0$};
\draw[->-=0.5] (3)--(0);
\draw [->-=0.35] (2)--(0);
\draw[bottom color=white] (-4,2)--(3.2,3.2)--(-.4,6)--cycle;
\node at (0,4) {Spatial Slice};
\node at (0,-2) {Triangulation of part of $CM$,
 with the cone point at time past infinity.};
\end{tikzpicture}
\end{center}

The ground state in the basis $|\Phi\rangle$ is
\begin{equation}\label{groundstate}
|\omega\rangle = \sum_{\Phi} \exp{\big[2\pi i\int_{CM} \omega(d\Phi)\big]} |\Phi\rangle.
\end{equation}
The integral over $CM$ indicates a sum over all simplices as above. The integrand is formed from a cocycle representative $\omega$ of the class of the action (note different representative give different states in the same phase). This $d$-cocycle is evaluated on each simplex using the labels given by $\phi$. The notation $d\Phi$ indicates the induced map $CM\to B\cG$. It means that we first extend $\Phi$ by the identity on all labels on the interior of $CM$. Then $d\Phi$ is the gauge field generated by a gauge transformation parametrized by this extension. Finally, $\omega$ is evaluated on this gauge field in the ordinary way to obtain an element of $\RR/\ZZ$. We draw it below for the figure above using the rules derived in section 4.

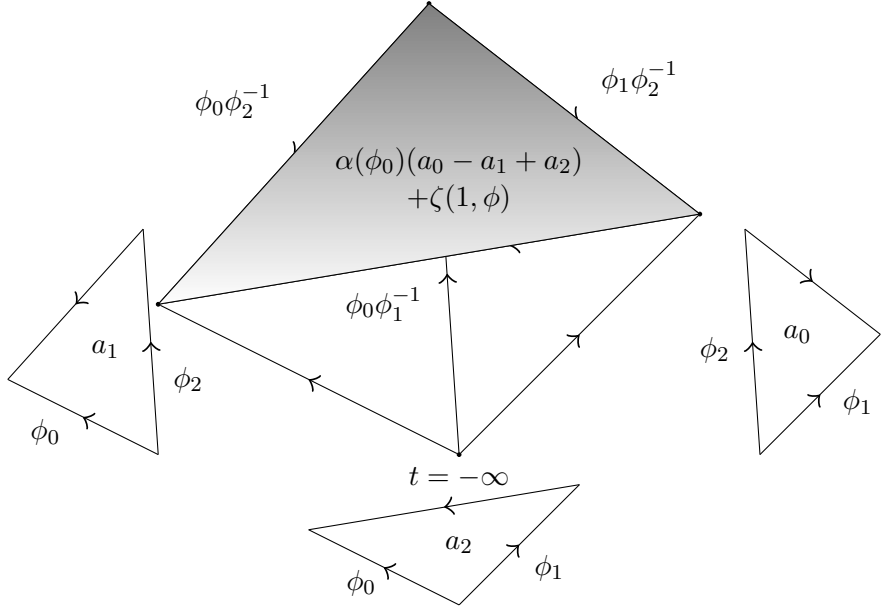
\begin{center}
\begin{tikzpicture}
[scale=1, vertices/.style={draw, fill=black, circle, inner sep=0.5pt}]
\node[vertices, label=below:{$t = -\infty$}] (1) at (0,0) {};
\node[vertices, label=right:{}] (2) at (3.2,3.2) {};
\node[vertices, label=left:{}] (0) at (-4,2) {};
\node[vertices, label=above:{}] (3) at (-0.4,6) {};
\node at (-1,2) {$\phi_0\phi_1^{-1}$};

\draw[->-=0.5] (1)--(2) ;
\draw[->-=0.5] (1)--(0) ;
\draw[->-=0.4] (1)--(3) node at (-3,4.7) {$\phi_0\phi_2^{-1}$};
\draw[->-=0.55] (3)--(2) node at (2.4,5) {$\phi_1\phi_2^{-1}$};
\draw[->-=0.5] (3)--(0);
\draw [->-=0.35] (2)--(0);
\draw[bottom color=white] (-4,2)--(3.2,3.2)--(-.4,6)--cycle;

\draw[->-=0.5] (4,0)--(5.6,1.6) ;
\draw[->-=0.5] (4,0)--(3.8,3) ;
\draw[->-=0.5] (3.8,3)--(5.6,1.6) ;

\draw[->-=0.5] (-4,0)--(-6,1) ;
\draw[->-=0.5] (-4,0)--(-4.2,3) ;
\draw[->-=0.5] (-4.2,3)--(-6,1) ;

\draw[->-=0.5] (1.6,-.4)--(-2,-1) ;
\draw[->-=0.5] (0,-2)--(-2,-1) ;
\draw[->-=0.5] (0,-2)--(1.6,-.4) ;

\node at (1.2,-1.5) {$\phi_1$};
\node at (-1.3,-1.7) {$\phi_0$};
\node at (0,-1.2) {$a_2$};
\node at (-4.7,1.4) {$a_1$};
\node at (-3.6,1) {$\phi_2$};
\node at (-5.5,0.3) {$\phi_0$};
\node at (4.5,1.6) {$a_0$};
\node at (3.4,1.4) {$\phi_2$};
\node at (5.3,0.7) {$\phi_1$};
\node at (0,3.9) {$\alpha(\phi_0)(a_0-a_1+a_2)$};
\node at (0,3.4) {$+\zeta(1,\phi)$};
\end{tikzpicture}
\end{center}

Note that since $d\Phi$ is a gauge transformation, $\omega(d\Phi)$ is exact, and so $\int_{CM}\omega(d\Phi)$ reduces to an integral over the boundary $\partial CM = M$, ie. the spatial slice, so \eqref{groundstate} is of a very similar form to the states considered in \cite{SPT}. In particular, it is short-range entangled in the sense of \cite{SPT}.

Let us derive the action of the 2-group $\cG$ on the matter fields. When the matter field configuration $\Phi=(a,\phi)$ is trivial, a parameter of the global $\cG$ symmetry is the same as a parameter $(\lambda,f)$ of a gauge symmetry except that $f$ is constant and $d\lambda = 0$. A global symmetry transformation $(\lambda,f)$ should act on the matter fields $\Phi\mapsto\Phi'$ so that $d\Phi' = d\Phi^{\phi\cdot \lambda,f}$, where on the right-hand side we have a gauge transformation with a parameter $(\phi\cdot \lambda,f)$. This will ensure that the state $|\omega\rangle$ is invariant (see below). Let us use this property to deduce $\Phi'$.

Let $\Phi=(a,\phi)$, then $d\Phi=(B,A)$, where
\begin{equation}\label{phi2form}
B=\zeta(1,\phi) + \phi \cdot da, \quad A=1^\phi.
\end{equation}
If we transform this field configuration by $(\phi\cdot \lambda,f)$, the 2-form part becomes
$$
f \cdot \zeta(1,\phi) + (f \phi)  \cdot d\lambda + \zeta(1^\phi, f) + (f \phi) \cdot da.
$$
Since $d\lambda = 0$, the second term vanishes, and by \eqref{zetanorm} and since $f$ is constant, so does the third. We are left with
$$
f \cdot( \zeta(1,\phi) + \phi \cdot da).
$$
Compare with \eqref{phi2form}. Meanwhile the 1-form part of $d\Phi'$ is simply $1^{f\phi}$. 

One can easily check that the same 2-connection can be obtained by acting on a trivial 2-connection by a gauge transformation parametrized by $\Phi' = (a+ \phi \cdot \lambda + \kappa(f,\phi), f\phi)$, where $\kappa$ was defined in (\ref{kappa}). In other words, the global symmetry transformation rules for matter fields are
$$
\phi \mapsto f \phi,
$$
$$
a \mapsto a + \phi\cdot\lambda + \kappa(f,\phi).
$$
Further, since $\omega$ is invariant under the simultaneous action of $\pi_1$ on itself and on $\pi_2$, the global symmetry transformation preserves $\omega(d\Phi)$ and therefore simply permutes the summands in $|\omega\rangle$, leaving the state invariant. Gauging the symmetry leads to the 2-group TQFT with cocycle $\omega$.

\subsection{Examples}

Let us collect here a couple examples of a interesting 2-groups. First is the 2-group with $\Pi_1 = (\ZZ/p)^3$, $\Pi_2 = \ZZ/p$, $\alpha$ trivial, and $\beta(a_1,a_2,a_3) = a_1 a_2 a_3$. A crossed module that realizes this 2-group is
$$
\ZZ/p \to UT(p,3) \to UT(p,4) \to \ZZ/p^3,
$$
where $UT(p,k)$ is the group of upper triangular kxk matrices over $\mathbb{F}_p$ with 1s on the diagonal (generalizing the Heisenberg group $k=3$). The center map places the three northeast elements of a 3x3 matrix into the three northeast elements of a 4x4 matrix.

We have already computed
$$
\zeta(a,f) = f_1 (a_2+df_2) (a_3+df_3) + a_1 f_2 (a_3+df_3) + a_1 a_2 f_3.
$$
and
$$
\kappa(g,\phi) = - g_1 \phi_2 d\phi_3.
$$
Under a global symmetry transformation parametrized by $(g,\lambda)$ we therefore have
$$
\phi_1 \mapsto \phi_1 + g_1
$$
$$
\phi_2 \mapsto \phi_2 + g_2
$$
$$
\phi_3 \mapsto \phi_3 + g_3
$$
$$
\alpha \mapsto \alpha + \lambda - g_1 \phi_2 d\phi_3.
$$

Our next example is a 2-group with $\Pi_1 = \ZZ/p$, $\Pi_2 = \ZZ/p$, $\alpha$ trivial, and
$$
\beta(a) = \widetilde{a} \frac{\delta \widetilde{a}}{p},
$$
where $\widetilde{a}$ denotes a lift of $a$, which is ordinarily just defined mod $p$, to an integer-valued cochain. Thus $\beta$ is an integer defined mod $p$. We have used this lifting throughout the paper so far, where we might write $\beta = a\delta a/p$, but it will be important for trivializing $\beta$ that we keep it explicit here. The derivation is a little bit technical, so one can feel free to skip to the global symmetry transformations.

Indeed, plugging in a gauge trivial configuration $a = \delta f$, the abridged notation makes $\beta$ look identically zero since $\delta^2 = 0$. However, $\delta \widetilde{\delta f} \neq 0$. Below we illustrate for $p=2$ how an $f$ supported at a vertex with both an incoming and outgoing edge has this property.

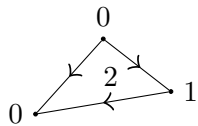
\begin{center}
\begin{tikzpicture}
[scale=1, vertices/.style={draw, fill=black, circle, inner sep=0.5pt}]
\node[vertices, label=right:{$1$}] (1) at (0.8,0.8) {};
\node[vertices, label=left:{$0$}] (0) at (-1,0.5) {};
\node[vertices, label=above:{$0$}] (2) at (-0.1,1.5) {};
\draw[->-=0.55] (2)--(1) node at (0.6,1.4) {};
\draw[->-=0.5] (2)--(0) node at (-0.75,1.35) {};
\draw [->-=0.5] (1)--(0) node at (0,1) {$2$};
\node at (0,-1) {An $f$ with $\delta \widetilde{\delta f} \neq 0$.};
\end{tikzpicture}
\end{center}

On the other hand, $\widetilde{\delta f} = \delta \widetilde f \mod p$, so since $\beta$ is defined mod $p$ we can write
$$
\beta(\delta f) = \delta \widetilde f \frac{\delta\widetilde{\delta f}}{p} = \delta ( \widetilde f \frac{\delta\widetilde{\delta f}}{p}).
$$
Call the potential in parenthesis on the right hand side $\zeta(0,f)$. Next we consider for constant $g$
$$
\zeta(0,f) - \zeta(0,f+g) = - \widetilde g \frac{\delta\widetilde{\delta f}}{p} = -\delta(  \widetilde g B(f)),
$$
where $B(f)$ is an integral 1-cochain made by labelling each edge with a $1$ if the value of $\tilde f$ at the source of the edge is larger than the value at the end of the edge. Let us show this identity. When forming $\widetilde{\delta f}$, we use an integral lift with values in $[0,p-1]$. Then the edges to which we must add $p$ to make positive are the ones that are negative. Since $\delta f$ on an edge $0\to 1$ is $f_1 - f_0$, this occurs iff $f_0 > f_1$. When computing $\delta \widetilde{\delta f}$, the values of $f$ all cancel and we are just left with these $p$s, which we divide by $p$ to get 1s. Putting this together, $\delta \widetilde{\delta f}/p = \delta B(f)$. The descendant potential we defined above for this example is thus $\kappa(g,f) = -\widetilde g B(f)$.

In the above description of the ground state of an SPT with this symmetry group, we have matter consisting of $\ZZ/p$ labels $\phi$ at vertices and $\ZZ/p$ labels $\alpha$ along edges. Under a global symmetry parametrized by $(g,\lambda)$ we have
$$
\phi \mapsto \phi + g
$$
$$
\alpha \mapsto \alpha + \lambda -  \widetilde g B(\phi).
$$
Gauging this symmetry produces a gauge field with gauge group given by the 2-group just described.

\section*{Appendix: Classifying space of a 2-group}

In this section we implicitly think of $\cG$ as a 2-category.

The classifying space of a 2-group $\cG$ is a topological space $B\cG$ with $\cG$ as its homotopy type. We can construct $B\cG$ inductively as a cell complex with 1 0-cell, 1-cells corresponding to 1-morphisms in $\cG$, 2-cells corresponding to 2-morphisms, 3-cells corresponding to relations among the 2-morphisms, 4-cells added to kill any $\pi_3$ introduced in the previous stage, 5-cells added to kill any $\pi_4$, and so on. Note that the relations among 1-morphisms are imposed by inserting the identity 2-morphisms, and any $\pi_2$ created among these is killed by the 3-cells imposing the relations among 2-morphisms.

Consider a configuration for the $\cG$ gauge theory. Using a section $s:\Pi_1 \to G$ to make the $A$ variables live in $G$ and including the $B$ variables in $H$, we can interpret this configuration as a composable diagram in $\cG$. Mapping each 0-cell of $X$ to the unique 0-cell of $B\cG$, 1-cells to the 1-cell of the corresponding 1-morphism (an element of $G$), and 2-cells to the 2-cell of the corresponding 2-morphism (this is generally an element of $H$ but the constraint \eqref{flatness} implies it is actually an element of $\Pi_2$), we obtain a map from the 2-skeleton of $X$ to $B\cG$. The 3-cell constraint \eqref{2-flatness} implies that we can extend this map to all of $X$. Changing which section one uses amounts to a gauge transformation of the original configuration.

Conversely, by cellular approximation any map $X\to B\cG$ gives us a configuration for the $\cG$ gauge theory. Cellular homotopies are gauge transformations, so we always get gauge-equivalent configurations if we pick  a different cellular approximation.

There is a map of 2-groups, which may be thought of as a functor, from $\cG$ to the group $\Pi_1$ (considered as a 2-group with only identity 2-morphisms) given by identifying isomorphic 1-morphisms in $\cG$. This gives a map on the 3-skeleton of $B\cG$ to $B\Pi_1$. Since the higher cells are added to kill homotopy groups for each space, we can inductively extend this to a map $B\cG \to B\Pi_1$. The fiber of this map over the unique 0-cell of $B\Pi_1$ is the classifying space $B^2\Pi_2$ of the group of 2-morphisms from the identity 1-morphism to itself. This space has second homotopy group $\Pi_2$ and all others vanishing.

This map is well known to be a fibration
\begin{center}
\begin{tikzpicture}
\node (BG) {$B\cG$};
\node (Bpi1) [below of=BG, node distance=1.1cm] {$B\Pi_1$};
\node(X) [left of=BG, node distance=1.5cm] {$B^2\Pi_2$};
\draw[->] (BG) to node {} (Bpi1);
\draw[->] (X) to node {} (BG);
\end{tikzpicture},
\end{center}
which is classified by the Postnikov class $\beta \in H^3(B\Pi_1,\Pi_2)$. For this fibration, the $E^2$ page of the Serre spectral sequence is the $\alpha$-equivariant cohomology $H^p(B\Pi_1,H^q(B^2\Pi_2,\mathbb{Z}))$. The shape of the relevant piece is
\begin{center}
\begin{tikzpicture}
\node (00) {$\mathbb{Z}$};
\node (01) [above of=00] {$0$};
\node (02) [above of=01] {$0$};
\node (03) [above of=02] {$\star$};
\node (04) [above of =03] {$0$};
\node (05) [above of=04] {$\star$};
\node (10) [right=1.2cm of 00] {0};
\node (11) [above of=10] {0};
\node (12) [above of=11] {0};
\node (13) [above of=12] {$\star$};
\node (14) [above of=13] {0};
\node (20) [right=1.2cm of 10] {$\star$};
\node (21) [above of=20] {0};
\node (22) [above of=21] {0};
\node (23) [above of=22] {$\star$};
\node (24) [above of=23] {0};
\node (30) [right=1.2cm of 20] {$\star$};
\node (31) [above of=30] {0};
\node (32) [above of=31] {0};
\node (33) [above of=32] {$\star$};
\node (40) [right=1.2cm of 30] {$\star$};
\node (41) [above of=40] {0};
\node (42) [above of=41] {0};
\node (50) [right=1.2cm of 40] {$\star$};
\node (51) [above of=50] {0};
\node (60) [right=1.2cm of 50] {$\star$};
\end{tikzpicture}.
\end{center}
Note that $p$ labels the columns and $q$ labels the rows.

The bottom row is $H^p(B\Pi_1,\mathbb{Z})$.

The rows with $q=1,2,4$ all vanish because Hurewicz's theorem implies $H_1(B^2\Pi_2,\mathbb{Z})=H_3(B^2\Pi_2,\mathbb{Z})=0$. From the universal coefficient theorem it then follows $H^1(B^2\Pi_2,\mathbb{Z})=0$, and since all cohomology classes on $B^2\Pi_2$ in positive degree are $|\Pi_2|$-torsion, the 2nd and 4th cohomology groups also vanish.

The universal coefficient theorem also tells us that $H^3(B^2\Pi_2)=\Hom(\Pi_2,\RZ)=\hat\Pi_2$, so the $q=3$ row is $H^p(B\Pi_1,\hat\Pi_2)$, where $\Pi_1$ acts on $\hat\Pi_2$ via $\alpha$. For example, $H^0(B\Pi_1,\hat\Pi_2)$ is the subgroup of $\Pi_1$-invariant characters in $\hat\Pi_2$.

It is also known that $H^5(B^2\Pi_2,\mathbb{Z})=H^4(B^2\Pi_2,\RZ)$ is the group of quadratic functions $q:\Pi_2\to\RZ$ \cite{EM}. The isomorphism is discussed in detail in \cite{KT1}. The group in the $(0,5)$ spot in the top left is then the subgroup of $\Pi_1$-invariant quadratic forms.

The first possibly non-zero differential is on the $E^3$ page:
$$
H^0(B\Pi_1,H^5(B^2\Pi_2,\ZZ))\to H^3(B\Pi_1,\hat\Pi_2).
$$
We find it difficult to prove, but we believe that this map sends an element of the left-hand side, which is a $\Pi_1$-invariant quadratic form $q:\Pi_2\to\RZ$ to $\langle \beta, - \rangle_q$, where the bracket denotes the bilinear pairing $\langle x, y \rangle_q = q(x+y)-q(x)-q(y)$.

The next possibly non-zero differentials are on the $E^4$ page:
$$
H^j(B\Pi_1,\hat\Pi_2)\to H^{j+4}(B\Pi_1,\ZZ)\simeq H^{j+3}(B\Pi_1,\RZ).
$$
This map is contraction with $\beta$.

The last relevant possibly non-zero differential is on the $E^6$ page:
$$
H^0(B\Pi_1,H^5(B^2\Pi_2,\ZZ))\to H^6(B\Pi_1,\ZZ).
$$
We believe this differential is actually zero.

This is enough to give the description of the topological actions we give in section 6.

\end{document}